\documentclass[10pt,twocolumn]{paper}
\usepackage{epsfig, graphicx}
\usepackage[english]{babel}
\usepackage[margin=1.5cm]{geometry}

\title{\center \rm \bf Adsorption of Polylysine on the Surface of the DMPS Monolayer}

\author{\small \rm   Yurii A. Ermakova$^{a}$, Victor E. Asadchikov$^{b}$,
Yurii O. Volkov$^{b}$, Alexander D. Nuzhdin$^{b}$, Boris S. Roshchin$^{b}$,\\
\small \rm Veijo Honkim\"aki$^c$, and Aleksey M. Tikhonov$^{d}$\/\thanks{tikhonov@kapitza.ras.ru}\\
\small$^a$ Frumkin Institute of Physical Chemistry and Electrochemistry, Russian Academy of Sciences, Moscow, 119071 Russia\\
\small$^b$ Shubnikov Institute of Crystallography, Federal Research Center Crystallography and Photonics,
Russian Academy of Sciences, \\
\small Moscow, 119333 Russia\\
\small$^c$ European Synchrotron Radiation Facility, 38000 Grenoble, France\\
\small$^d$ Kapitza Institute for Physical Problems, Russian Academy of Sciences, Moscow, 119334 Russia
}

\begin{document}
\maketitle

\abstract{ \it \normalsize
The effect of the adsorption of a polypeptide on the lateral interaction of dimyristoylphosphatidylserine molecules in different phase states on the surface of a 10\,mM KCl aqueous solution has been studied. Changes in the surface pressure and Volta potential induced by the adsorption of large poly-D-lysine molecules (about 200 links in a chain) have been determined at different areas per lipid molecule in a monolayer. The adsorption of macromolecules noticeably increases the elasticity of the monolayer under lateral compression in the liquid expanded state of lipid and reduces the effective dipole moment from 0.48 to 0.38 D. These properties are in qualitative agreement with X-ray reflectometry data for the lipid monolayer obtained with synchrotron radiation with a photon energy of $\approx 70$\,keV. The electron density profiles perpendicular to the surface of the aqueous subphase have been reconstructed from reflectometry data within a model approach to the structure of an interface with two and three layers. These profiles indicate the existence of a wide diffuse polymer layer $(150\pm40)$\,{\AA} in width at the interface of the monolayer in both the liquid expanded and liquid condensed states. A decrease in the area per molecule in the monolayer by a factor of 2 results in the doubling of the surface density of the macromolecule film. The adsorption of the polymer also affects the integral density of the layer of polar phospholipid groups, which decreases by a factor of $\approx 2$ in the liquid expanded phase and by $\sim 30$ \% in the liquid condensed phase.

}

\vspace{0.25in}
\normalsize

{\bf INTRODUCTION}

Lipid models of biomembranes have been widely used for a long time to study structural factors in the
interaction of various membrane active compounds with the surface of cells \cite{Mohwald,Stefaniu,Cevc}. Special attention is focused on the effect of various polypeptide macromolecules, which were applied in various biomedical applications, on the structure of the lipid matrix of cell walls. The interpretation of experimental data usually includes an important fundamental property of saturated phospholipids to demonstrate a first-order phase transition from a "liquid expanded" (LE) state to a "liquid condensed" (LC) one under the variation of the temperature $T$ or the lateral pressure $\Pi$ \cite{Loesche}. In the general case, in the LÑ monolayer state, diverse lipid mesophases both crystal and hexatic are observed; they differ between each other in, e.g., area per molecule $A$ and the angle of canting of aliphatic tails with respect to the normal to the surface \cite{Kaganer, Shih, Peterson, Helm, Kenn, Meijere}. We previously showed that change in the lateral packing of negatively charged phosphatidylserine molecules in model lipid membranes is also initiated by the adsorption of multivalent cations and large polypeptide molecules on their surface \cite{Erm1,Erm2,Mar1}. A good signature of this phenomenon appeared to be change in the dipole component of the boundary potential \cite{Erm3}. It can be expected that the clustering of lipids in the presence of polypeptide and the appearance of the dipole component of the boundary potential have many features of the molecular nature of the LE-LC phase transition.

In this work, to study the effect of the adsorption of polypeptide on the lateral interaction of lipid molecules in different phase states, we use dimyristoylphosphatidylserine (DMPS), which is saturated analog of phosphatidylserine. We report experimental data indicating structural changes in the DMPS monolayer in both the LE and the LC state in the presence of poly-D-lysine hydrobromide molecules at its water interface. These data were obtained when studying the compressibility of the lipid monolayer using the Langmuir monolayer technique and from X-ray reflectometry with synchrotron radiation.

\vspace{0.15in}
{\bf EXPERIMENT}

C$_{34}$H$_{65}$NO$_{10}$PNa sodium salt of DMPS phospholipid (Avanti Polar Lipids) was prepared at a concentration of $\approx 0.5$\,mg/mL in 5 : 1 volume mixed chloroform–methanol solutions (Merck KGaA and Macron Fine Chemicals, respectively). As a subphase, we used $10$\,mM solutions of KCl in triply distilled water (ðÍ = 6–7), including those with the addition of [H$_{12}$C$_6$N$_2$O]$_n\times$HBr poly-D-lysine hydrobromide, where $n \approx 200$ (P7886, Sigma-Aldrich). The concentration of polylysine added to the subphase was $ñ = 0.0134$\,mg/mL ($\sim 0.27$\,$\mu$M. The DMPS monolayer was deposited by a Hamilton syringe on the surface of the aqueous solution with the necessary amount of lipid, which corresponded to areas of 36\,\AA$^2$ and 72\,\AA$^2$ per molecule in the monolayer at the measurement of X-ray scattering. Pressure–area diagrams and the Volta potential were measured simultaneously at a MicroTrough XS, V.4.0 setup (Kibron Inc., Finland) with a $20\times6$\,cm tetrafluoroethylene bath, two polyoxymethylene barriers, and a vibrating Kelvin electrode to detect changes in the boundary potential. Before each experiment, in the absence of the lipid monolayer, we performed test measurements of the surface pressure (tension) in the aqueous solution of polypeptide. These measurements revealed no surface activity at the maximum approach of barriers. All measurements were performed at room temperature $\sim20$\,min after the deposition of lipid and at a compression rate of $\sim10$\,mm$^2$/min. The typical reproducibility of compression diagrams was no worse than 1-3\%.

The compression diagrams of DMPS monolayers shown in Fig. 1a reproduce well the results reported in
our previous works \cite{Erm1,Asad,Erm4}. Both curves clearly demonstrate a section of a smooth increase in the lateral pressure in the region of the LE state of the monolayer. According to previous studies and molecular dynamics simulations of such systems, the aqueous surface in this region includes lipid domains whose total area increases as barriers approach each other and the lateral pressure increases. This region is followed by a small plateau with an intermediate state and a sharper increase in the pressure at the transition to the LC state with the maximally dense packing of lipids near collapse. Thus, the compression of the lipid monolayer successively changes its physical state beginning with a "two-dimensional gas" at pressures below 1–2\,mN/m to the LE and LC states and, finally, to the collapse of the monolayer when lipids go out the bath or/and from the monolayer to the aqueous medium.

\begin{figure}
\vspace{0.5in}
\epsfig{file=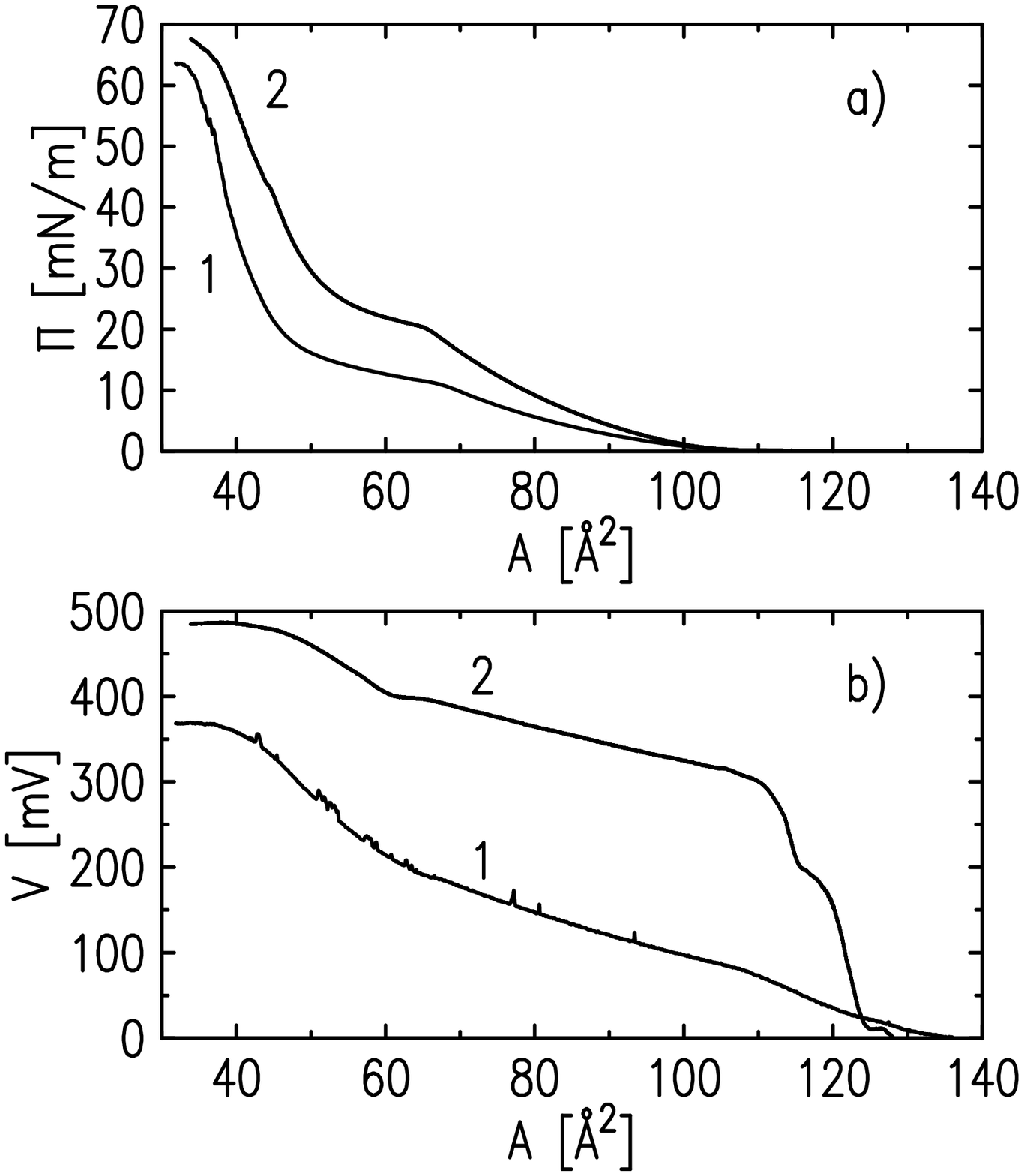, width=0.4 \textwidth}

\vspace{0.15in}
\small {\bf Figure 1.} \it (a) Compression diagrams and (b) Volta potential of the DMPS monolayer measured at the deposition of lipid on the surface of the 10\,mM KCl aqueous solution in the (1) absence and (2) presence of poly-D-lysine molecules with the concentration $c = 0.0134$\,mg/mL in this solution.
\end{figure}

The data presented in Fig. 1b make it possible to identify three segments of the variation of the Volta potential whose positions correlate with the $\Pi - A$ diagram, where $\Pi$ is the surface pressure and $A$ is the area per molecule. The potential in the LE region varies according to an almost linear law, whereas the poten tial in the LC region increases more sharply to the plateau, which characterizes the maximally dense monolayer immediately before the collapse. All phases listed above are manifested in the variation of the edge Volta potential. In a very strongly diluted monolayer, a Kelvin vibrating electrode detects random lipid "spots," which are located under its surface, whereas the surface tension of this surface remains as a whole unchanged and close to the surface tension of pure water $\gamma_0$ (under normal conditions, $\gamma_0 \approx 72.5$\,mN/m). In this region, random fluctuations of the electric potential are detected. In the case of the formation of an almost continuous monolayer, possibly with the inclusion of a few water spots, the surface tension changes noticeably, the pressure becomes above 1\,mN/m, and the variations of the pressure and potential beginning with this time reflect the lateral interaction of lipid molecules.

\begin{figure}
\vspace{0.5in}
\epsfig{file=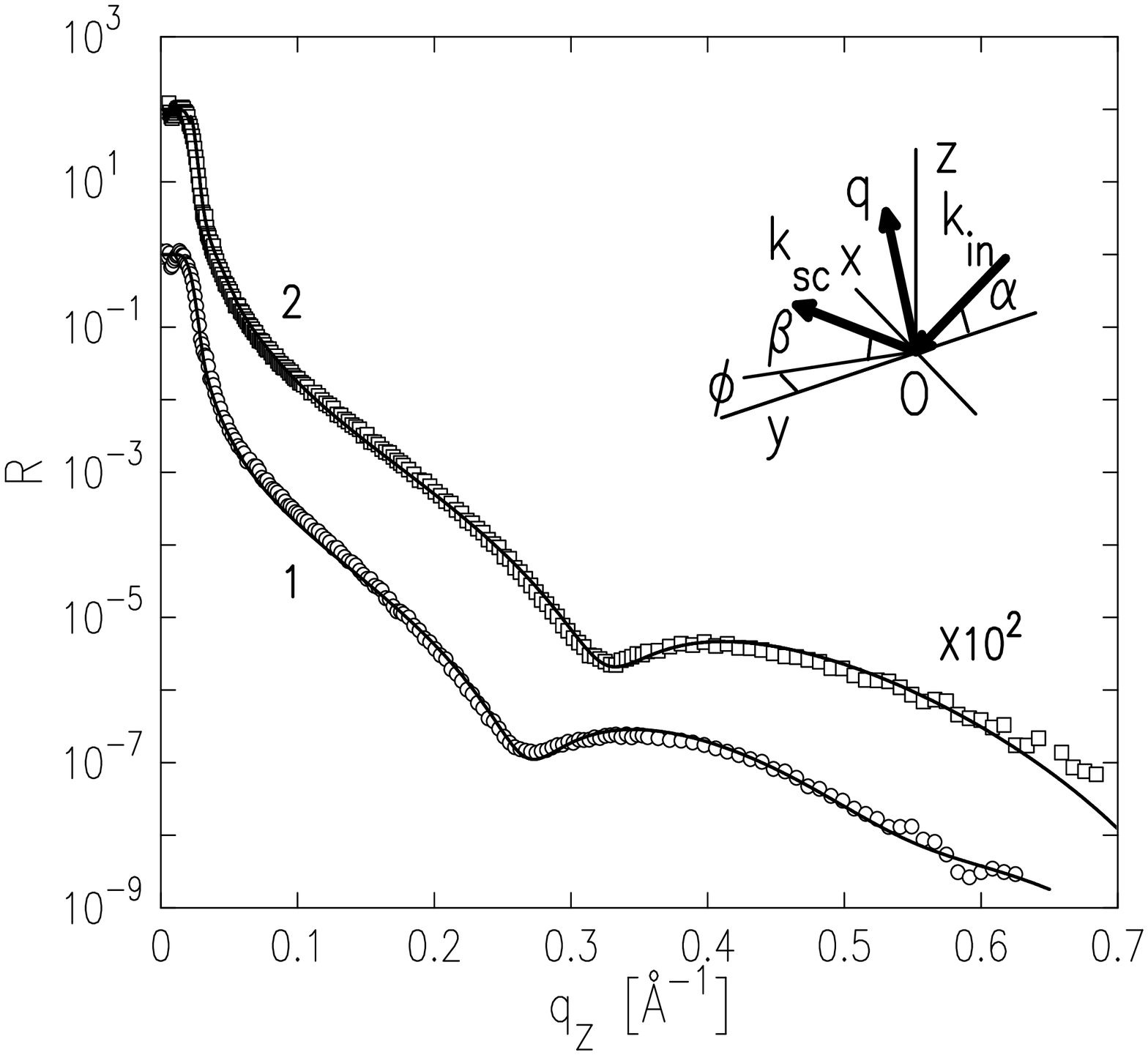, width=0.4 \textwidth}

\vspace{0.15in}
\small {\bf Figure 2.} \it Dependence $R(q_z)$ for the DMPS monolayer on the surface of water at an area of 72\,{\AA}$^2$ for the lipid monolayer (circles) without and (squares) with the adsorption polymer layer. Lines 1 and 2 are the respective model calculations. The inset shows the kinematics of scattering in the coordinate system where the $xy$ plane coincides with the interface between the monolayer and water, the $Ox$ axis is perpendicular to the beam direction, and the $Oz$ axis is opposite to the gravitational force perpendicular to the surface; {\bf k}$_{\rm in}$ and {\bf k}$_{\rm sc}$ are the wave vectors of the incident and scattered beams, respectively; correspondingly, {\bf q = k$_{\rm in}$ {\rm -} k$_{\rm sc}$} is the scattering vector.
\end{figure}

The reflection coefficient of X rays, $R$, for Langmuir DMPS monolayers at the water–air interface was measured on the ID31 beamline of the European Synchrotron Radiation Facility (Grenoble, France) \cite{ID31}. The intensity of the focused monochromatic photon beam with a wavelength of $\lambda\approx0.18$\,\AA  (the photon energy $E\approx70$\,keV, $\Delta E/E =0.4\%$) and with the cross section $\sim 6$\,$\mu$m in height and $\sim 150$\,$\mu$m in horizontal plane was $\sim 10^{10}$\,photons/s. The method of measurement of the reflection coefficient $R$ was described in \cite{Bitto,Malkova,Koo,Tikh22}. Monolayer samples of DMPS phospholipid were prepared under conditions similar to those at the measurement of compression diagrams but in a circular tetrafluoroethylene bath 10\,cm in diameter, which was placed in a hermetic thermostat with X-ray transparent windows.

\begin{figure}
\vspace{0.5in}
\epsfig{file=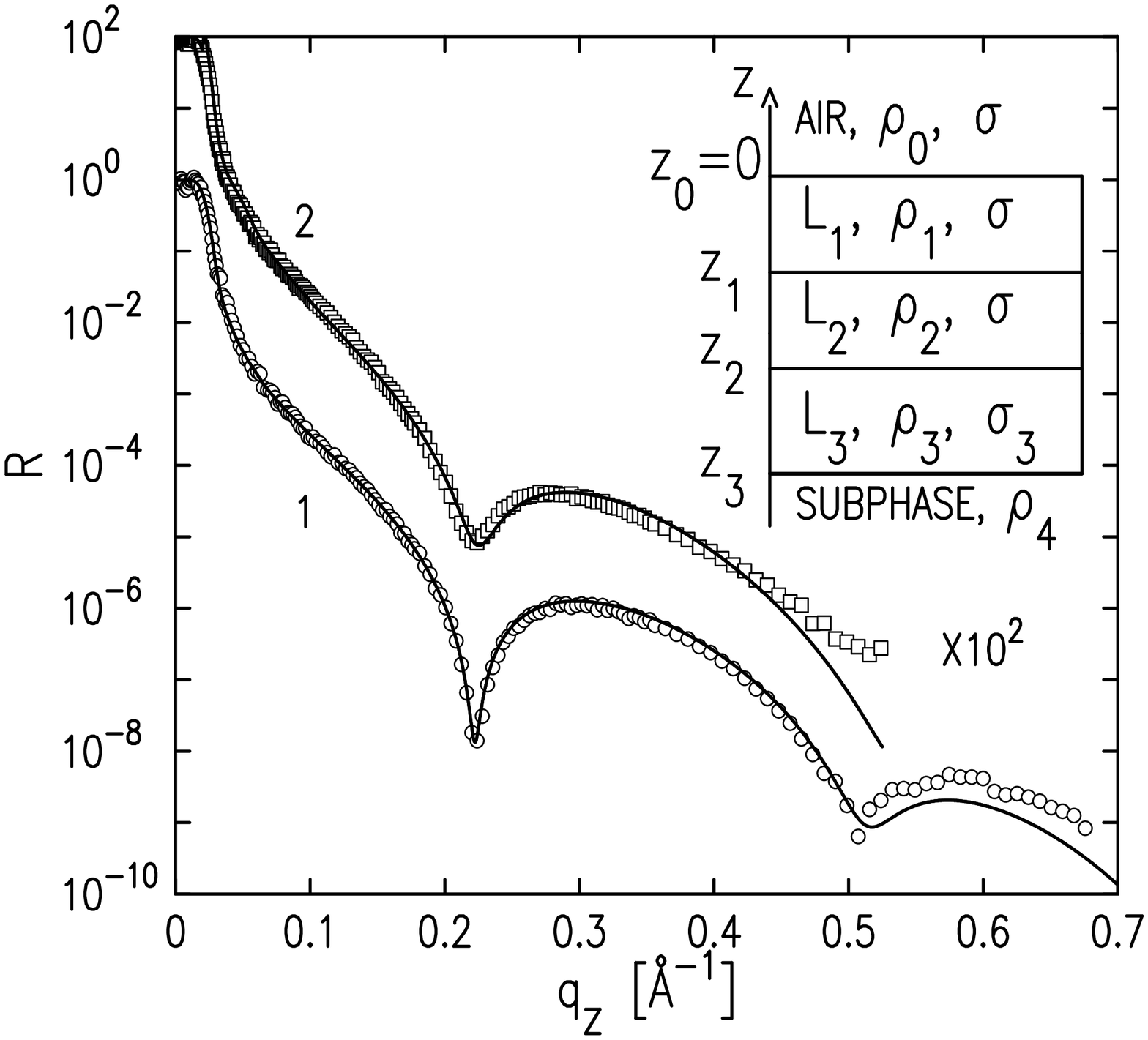, width=0.4 \textwidth}

\vspace{0.15in}
\small {\bf Figure 3.} \it Dependence $R(q_z)$ for the DMPS monolayer on the surface of water at an area of 36\,{\AA}$^2$ for the lipid monolayer (circles) without and (squares) with the adsorption polymer layer. Lines 1 and 2 are the respective model calculations. The inset shows the parameterization of the structure of the air air–water interface.
\end{figure}

\vspace{0.15in} 
{\bf THEORY}

In the case of specular reflection, the scattering vector {\bf q} has only one nonzero component $q_z=(4\pi/\lambda)\sin\alpha$, where $\alpha$ is the glancing angle in the plane normal to the surface (see the inset of Fig. 2). The total external reflection angle for the surface of water $\alpha_c\approx\lambda\sqrt{r_e\rho_w/\pi}$$\approx0.018^\circ$ ($q_c=(4\pi/\lambda)\sin\alpha_c$ $\approx0.022$\,\AA$^{-1}$) is specified by the electron density in it $\rho_w\approx 0.333$  {\it e$^-$/}{\AA}$^3$, where $r_e = 2.814\cdot10^{-5}$\,{\AA} is the classical electron radius. Figures 2 and 3 show the experimental dependences $R(q_z)$ obtained for two LC and LE surface states with $A=36$\,{\AA}$^2$ and $72$\,{\AA}$^2$, respectively. It is seen that the addition of the polymer to the subphase affects the shape of the reflection line, which indicates the formation of a lipid–polymer film.

The reflectometry data were analyzed within a model approach, where the structure of the film at the
air–water interface was divided into $N$ layers. Each layer has the thickness $L_j$ and the electron density $\rho_j$, and the widths of the interfaces in the multilayer are described by the parameters $\sigma_j$ (see the inset of Fig. 4). The electron density profile $\rho(z)$ in the surface structure along the normal to the interface is parameterized as \cite{Buff,Wu,Weeks}
\begin{equation}
\displaystyle
\rho=\frac{1}{2}\rho_{0}+\frac{1}{2}\sum_{j=0}^N(\rho_{j+1}-\rho_j)
{\rm erf}\left(\frac{z_j}{\sigma_j\sqrt{2}}\right),
\end{equation}
where ${\rm erf}(x)$ is the error function, $z_j=z+\sum_{n=0}^{j}L_n,$ ($L_0 = 0$), 
and $\rho_{0}=0$ and $\rho_{N+1}=\rho_w$ are the electron densities in air and in the bulk of water. In the distorted wave Born approximation (DWBA), the reflection coefficient $R(q_z)$ is expressed in terms of the parameters of electron density (1) as \cite{Sinha}
\begin{equation}
\begin{array}{c}
\displaystyle
R(q_z)\approx \left|\frac{\displaystyle q_z-q_z^t}{\displaystyle q_z+q_z^t}\right|^2
\\ \\
\displaystyle
\times
\left|\sum_{j=0}^{j=N}{\frac{\rho_{j+1}-\rho_{j}}{\rho_w}e^{\displaystyle -\left(q_zq_z^t\frac{\sigma^2_j}{2}-iz_j\sqrt{q_zq_z^t}\right)}}\right|^2.
\end{array}
\end{equation}
where $q_z^t=\sqrt{q_z^2 - q_c^2}$. In the general case, the desired structure is found by fitting Eq. (2) with $3N+1$ free parameters $\rho_j, l_j$, and $\sigma_j$ to experimental data for $R(q_z)$ .

\begin{figure}
\vspace{0.5in}
\epsfig{file=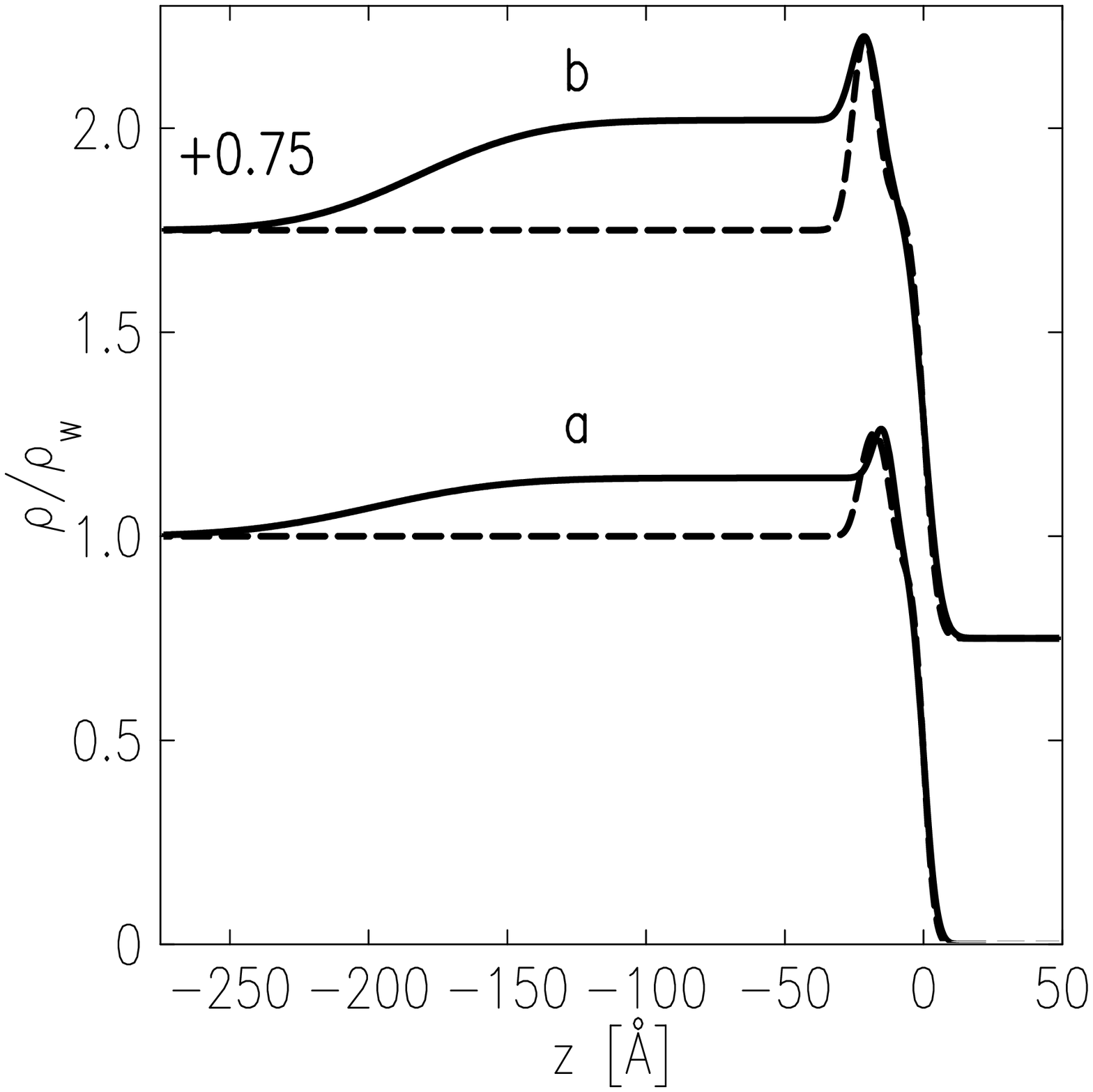, width=0.4 \textwidth}

\vspace{0.15in}
\small {\bf Figure 4.} \it Electron density profiles $\rho(z)$ in units of the electron density in water under normal condition $\rho_w\approx 0.333${\it e$^-$/}{\AA}$^3$ for the monolayer (dashed lines) without and (solid lines) with the polymer. The point $z = 0$ is chosen at the interface between lipid molecules with air. Lines $a$ and $b$ correspond to the liquid (LE, $A=72$\,{\AA}$^2$) and condensed (LC, $A=36$\,{\AA}$^2$) states of the monolayer. For convenient comparison, lines $b$ are shifted along the $y$ axis by 0.75.
\end{figure}

In the experiment, thermal fluctuations of the surface (capillary waves) in the illuminated region ($\sim 1$\,mm$^2$) make a contribution to the observed structure, which smears jumps in the electron density profile \cite{Braslau2}. In the calculations, the "capillary width" $\sigma^2 = k_BT/(2\pi\gamma(A))\ln(Q_{max}/Q_{min})$, where $k_B$ is the Boltzmann constant, $\gamma(A)=\gamma_0 - \Pi(A)$, $Q_{max} = 2\pi/a$ ($a\approx 10$\,\AA{} is the intermolecular distance) is the short wavelength limit in the capillary wave spectrum, and $Q_{min}=q_z^{max}\Delta\beta$ ($\Delta\beta \approx 0.023^\circ$ is the angular resolution of the detector) is the long-wavelength limit of surface fluctuations involved in the experiment, was used as the parameter $\sigma_j$ for the boundaries of the monolayer. We successfully used such an approach to analysis of reflectometry data, e.g., to study structures and phase transitions in adsorption amphiphilic films on macroscopically planar oil–water interfaces \cite{c22,PRL,Tikh21,Ping,TAMSCH}.

\vspace{0.15in}
{\bf RESULTS}

All experimental dependences $R(q_z)$ can be quite well described within two- and three-layer models.
The structure of the water–lipids–air interface can be conditionally represented in the form of two layers
($N=2$). In view of the chemical structure of the DMPS molecule, it is reasonable to suggest that the
first layer with the thickness $L_1 = 10\div15$\,\AA{} and electron density $\rho_1 = 0.92\div1.06\rho_w$ is formed by aliphatic segments of the molecule and its C$_{14}$H$_{27}$ tails. The second layer with the thickness $L_2\sim 10$\,\AA{} and electron density $\rho_2 = 1.2 \div 1.5\rho_w$ is formed by polar groups of phosphatidylserine. The parameter $\sigma = 3 \div 5 $\,\AA{}, whose calculation was described above, was taken as the widths $\sigma_0$, $\sigma_1$, and $\sigma_2$. There parameters of the DMPS monolayer in the LE and LC states are in agreement with \cite{Erm4}.

The adsorption of the polymer is clearly manifested in changes in the reflectivity curves for both states of the monolayer. This adsorption is described well by the three-layer model ($N=3$), where the packing
parameters $\rho_1$ and $L_1$ of aliphatic tails of the phospholipid are equal to the respective parameters of the pure monolayer in the corresponding state. Good agreement of the calculated curves with the experiment is reached by varying the parameters $\rho_2$ and $L_2$ of the layer of polar groups, the thickness $L_3$ and density $\rho_3$ of the polymer film, and the width $\sigma_3$ of the polymer film–subphase interface.

According to this procedure, the adsorption of the polymer first affects the integral density $\rho_2L_2$ of the layer of polar groups: this density decreases noticeably (by a factor of $\approx 2$) in the LE phase, whereas this decrease is much less ($\sim 30$\,\%) in the LC phase. Second, the fitting $\rho_3$ value depends on the density of the monolayer: the densities of the polymer film in the LE and LC states are $\rho_{3}^{à} \approx 1.14\rho_{w}$ and $\rho_{3}^{b} \approx 1.27 \rho_w$, respectively. The parameter $L_3 = 150 \pm 40$\,\AA{} is determined with a large error and weakly depends on the density of the monolayer. Consequently, a decrease in the area per molecule in the monolayer from 72\,{\AA} to 36\,{\AA} leads to an almost doubling of the surface density, $\rho_3$, of the adsorption film of macromolecules: ($(\rho_{3}^{b}-\rho_w)/(\rho_{3}^{a}-\rho_w) \approx 2$. Third, the width $\sigma_3 \approx 40$\,\AA{} is much larger than the capillary width $\sigma$, which indicates the diffuse structure of the polymer film–subphase interface. The calculated curves for $R(q_z)$ are shown by solid lines in Figs. 2 and 3 and the corresponding model profiles $\rho(z)$ are shown in Fig. 4.

\begin{figure}
\vspace{0.5in}
\epsfig{file=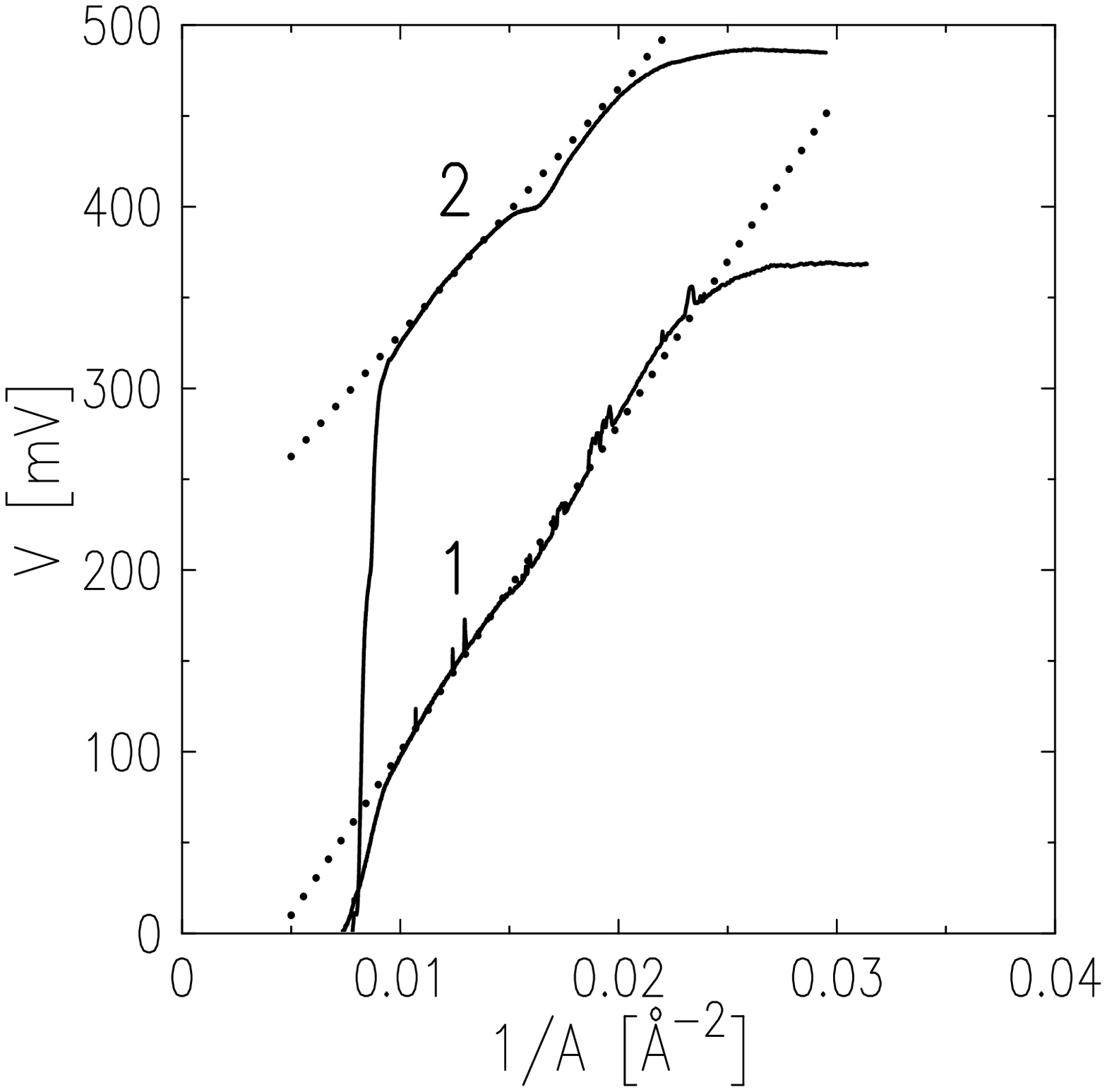, width=0.4 \textwidth}

\vspace{0.15in}
\small {\bf Figure 5.} \it (Solid lines) Volta potential of DMPS monolayers (see the notation in Fig. 1) versus the density of lipid in the monolayer on the surface of the corresponding solutions. Dotted straight lines are approximations of the linear segments of the curves. The effective dipole moments estimated from the Helmholtz relation on the linear segments of curves 1 and 2 are 0.48 and 0.38 D, respectively.
\end{figure}

\vspace{0.15in}
{\bf DISCUSSION}

Changes in the boundary potential of monolayers in the liquid crystal state described in many works can be attributed to a simple increase in the density of effective dipole moments of lipid molecules. These changes satisfy well the simple Helmholtz relation $\Delta V=12\pi\mu/A$ , where $V$ is the change in the boundary potential in millivolts, $\mu$ is the effective dipole moment in millidebyes, and $A$ is the area per molecule in angstroms squared \cite{Brockman}. This linear dependence is observed in the segments of curves shown in Fig. 1b that correspond to the LE state of the monolayer and are shown in Fig. 5.
The slope of this dependence becomes much smaller in the presence of high molecular polypeptide
in the solution. This means that the effective dipole moment in these segments of compression of the
monolayer decreases from 0.48 to 0.38\,D. This value is obviously the sum of all dipole moments that are
determined by the structure of the phospholipid and are due to the partial immersion of cations of the electrolyte in the polar region and to the orientation of water molecules associated with this region. For this reason, the molecular interpretation of the decrease in the effective dipole moment is a complex problem, which can be solved, e.g., by numerical molecular dynamics simulation of these systems.

According to the measurement and the compression diagram of DMPS monolayers on the surface of
KCl solutions in the presence of polylysine in it (Fig. 1à), the effect of poly-D-lysine is the most pronounced in the region of the LE state. A larger slope of experimental curves reflects an increase in the stiffness of the monolayer under compression. At the same time, the phase transition in the condensed state begins at the same area per lipid molecule below 70\,{\AA}$^2$ per molecule. The compression curve in the condensed state has a slope close to that of the "pure" DMPS monolayer, but is shifted toward larger areas per lipid molecule. This area at a lateral pressure of $\sim40$\,mN/m increases from 39 to 46\,{\AA}$^2$. The polymer most probably groups lipids to denser clusters. Indeed, if the polymer filled aqueous "cavities" in the monolayer, the compression curves would be shifted toward larger areas. Since polylysine in test experiments did not exhibit surface activity, the shift of the curves toward larger areas per molecule in the LC state cannot be attributed to the introduction of links of the polymer between lipid molecules, as is possible for hydrophobic molecules.

\vspace{0.15in}
{\bf CONCLUSION}

To summarize the above results, we note that the choice of the polypeptide, its density, and a sufficiently large molecular mass ensures the maximum expected filling of the surface by macromolecules. Furthermore, according to studies of the adsorption of polylysines on the surface of bilayer lipid membranes,
in particular, black lipid membranes and those in the suspension of liposomes, their adsorption occurs only on negatively charged surfaces and with a very high efficiency, filling the entire surface available for
adsorption \cite{Mar1,Mar2}. The character of the filling of the surface depends on the polymer–bilayer equilibrium constant. This constant for polylysines with a long chain estimated within a comparatively simple theoretical model is no more than $10^4$ Ì$^{–1}$, which means their irreversible coupling with phospholipids \cite{Mol}. However, it was found that the presence of a layer of polylysines does not affect the stability of membranes, in contrast to polymers with hydrophobic segments, which can be partially incorporated into a membrane and create pores penetrable for ions. These facts mean that the shift of the compression curves toward larger areas per molecule seen in Fig. 1 is due not to the introduction of the polymer into the structure of the monolayer but is due to an increase in its stiffness in the LE state. The compressibility of the monolayer in the LC state hardly depends on the presence of the polymer in the aqueous substrate. This conclusion is confirmed by X-ray reflectometry data in Figs. 2 and 3 (the shape of the reflection curve for the LC state of the monolayer at the adsorption of poly-D-lysine generally does not change) and by the reconstructed electron density profiles in Fig. 4. Changes in the reflectivity curves are the most significant in the region where the monolayer behaves as a two-dimensional
liquid. The structural parameters of the interface estimated within the two-layer model are in good agreement with the geometrical characteristics of the amphiphilic lipid molecule. We believe that the
described methods for the study of complex polymerlipid systems at the water–air interface are useful for
various charged macromolecules having biomedical applications. Corresponding experimental results will
allow confirming or rejecting molecular structures actively analyzed by molecular dynamics methods.

\vspace{0.15in}
{\bf ACKNOWLEDGEMENTS}

The experiments were performed on the ID31 beamline of the European Synchrotron Radiation Facility (Grenoble, France) within the SC-4845 project. We are grateful to Helena Isern and Florian Russello (ESRF) for assistance in the preparation of the experiments. This work was supported by the Ministry of Science and Higher Education of the Russian Federation (state assignments for the indicated institutions) and by the Russian Foundation for Basic Research (project no. 16-04-00556).

\end{document}